\documentclass[%
 reprint,
superscriptaddress,
 amsmath,amssymb,
 aps,
]{revtex4-2}

\usepackage{graphicx}
\usepackage{dcolumn}
\usepackage{bm}
\usepackage{hyperref}

\usepackage{amsmath}
\usepackage{amssymb}
\usepackage{comment}
\usepackage{graphicx}
\usepackage{ulem}
\usepackage{color}

\begin{document}


\title{Biologically relevant finite-size effects in a driven lattice gas with particle pausing and dynamical defects}

\author{Johannes Keisers}
\thanks{These authors contributed equally to this work}
\affiliation{%
 Centre de Biologie Structurale (CBS), Université de Montpellier, CNRS, INSERM, Montpellier, France
}
\author{Lorenzo Vito Dal Zovo}%
\thanks{These authors contributed equally to this work}
\affiliation{%
 Centre de Biologie Structurale (CBS), Université de Montpellier, CNRS, INSERM, Montpellier, France
}%
\affiliation{Dipartimento di Scienze Matematiche “G. L. Lagrange”, Politecnico di Torino, Corso Duca degli
Abruzzi, 24, 10129 Turin, Italy}

\author{Norbert Kern}
\affiliation{Laboratoire Charles Coulomb (L2C), Université de Montpellier, CNRS, Montpellier, France}

\author{Luca Ciandrini}
\email{luca.ciandrini@umontpellier.fr}
\affiliation{%
 Centre de Biologie Structurale (CBS), Université de Montpellier, CNRS, INSERM, Montpellier, France
}%
\affiliation{Institut Universitaire de France}

\date{\today}

\begin{abstract}
In this article we present a comprehensive study of the totally asymmetric simple exclusion process with pausing particles (pTASEP), a model initially introduced to describe RNAP dynamics during transcription. We extend previous mean-field approaches and demonstrate that the pTASEP is equivalent to the exclusion process with dynamical defects (ddTASEP), thus broadening the scope of our investigation to a larger class of problems related to transcription and translation.
We extend the mean-field theory to the open boundary case, revealing the system’s phase diagram and critical values of entry and exit rates. However, we identify a significant discrepancy between theory and simulations in a region of the parameter space, indicating severe finite-size effects.
To address this, we develop a single-cluster approximation that captures the relationship between current and lattice size, providing a more accurate representation of the system’s dynamics. 
Finally, we extend our approach to open boundary conditions, demonstrating its applicability in different scenarios. Our findings underscore the importance of considering finite-size effects, often overlooked in the literature, when modelling biological processes such as transcription and translation. 
\end{abstract}
\maketitle

\section{\label{sec:level1}Introduction}
\begin{figure}[!t]
    \includegraphics[width=0.8\columnwidth]{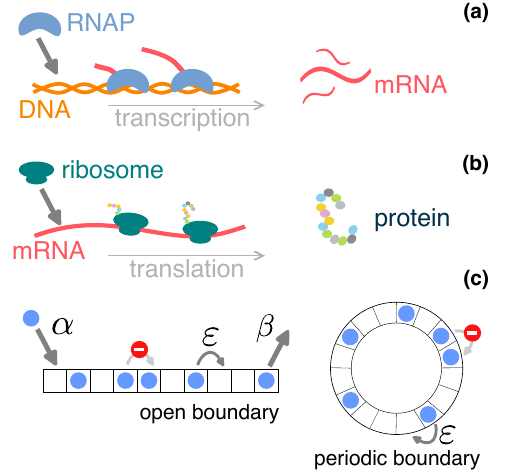}
\caption{\label{fig:transcription_translation_model} 
(a) RNA polymerases (RNAPs) bind to regions of the DNA and transcribe the genetic codes into messenger RNAs (mRNAs); (b) Ribosomes then translate the mRNAs into proteins. Both process are continuous in time but discrete in space: RNAPs advance on the DNA nucleotide by nucleotide to polymerase the mRNA molecule, while ribosomes move on mRNAs codon by codon (one codon is three nucleotides) to translate the genetic code from the language of nucleotides to amino-acids. (c) The TASEP is a traffic model that describes particles moving on a discrete lattice in a preferred direction. It can be formulated for open boundaries by considering particle injection and depletion, or for periodic boundaries, with fixed particle density.}  
\end{figure}

Protein synthesis is the biological process beginning with the {\it transcription} of the DNA into messenger RNA (mRNA) molecules, which are then {\it translated} from the language of nucleotides into assembled proteins made of amino acids~\cite{alberts2022molecular}. 
Both steps are carried out by molecular motors that advance, unidirectionally, on their template: RNA polymerases (RNAPs) proceed on genes and synthesise the mRNAs, that in turn serve as template for ribosomes, the molecular assembly factories that manufacture proteins starting from the nucleotide sequence (see Fig.~\ref{fig:transcription_translation_model}).

These biopolymerisation processes bear similarities with a driven lattice gas: The biological template (DNA or mRNA) is represented by a finite and discrete one dimensional lattice, and RNAP or ribosomes as particles hopping forward from site to site, synthesizing mRNAs or proteins along the way, as shown in Fig.~\ref{fig:transcription_translation_model}(c). As particles cannot overtake each other, queuing phenomena arise and the final biosynthesis current will  depend heavily on the particle (RNAP or ribosome) density on the template~\cite{Klumpp2008, ciandrini2013}.

It was precisely for modelling the kinetics of ribosomes on mRNAs that MacDonald and coworkers conceived in 1968~\cite{MacDonald1968} the totally asymmetric simple exclusion process (TASEP), a driven lattice gas model that we describe in a dedicated section. After their work, the exclusion process has been extensively studied and became a paradigmatic model in non equilibrium physics~\cite{Chou2011}. 
Since then, the TASEP and its variants has been extensively used to model the transcription and translation processes~\cite{Shaw2003, Klumpp2008, tripathi2008interacting, ciandrini2010, ciandrini2013, Wang2014, szavits2018deciphering, Waclaw2019, erdmann2020key}. 
Several variants of the model have been developed in order to improve its realism, some of them focusing on the particle's stepping cycle~\cite{basu2007traffic, klumpp2008effects, ciandrini2010, rousset2019exclusion}. In these variants, particles are taken to undergo transitions between different states, affecting their hopping kinetics, in an attempt to mimic the states of complex molecular machineries such as RNAPs and ribosomes.

In this article we study a particular extension of the exclusion process in which particles can stochastically enter end exit a `pausing' state. This variant has been introduced in~\cite{Klumpp2008} to model RNAP pausing during transcription. To further investigate this model in the context of mRNA translation, we are motivated by the existence of a class of antibiotics that disrupts protein synthesis by binding temporarily to ribosomes, blocking them for a stochastically determined period of time~\cite{tritton_ribosome-tetracycline_1977, harvey1980, wilson_ribosome-targeting_2014, savelsbergh_distinct_2009, stanley_structures_2010}.

Here we first extend the pausing TASEP to open boundary conditions, formerly limited to periodic systems. We explore different parameter regimes and remark that previous analytical approaches fail in a regime of rare and long pauses; we point out how this regime is dominated by finite-size effects, and develop a theoretical framework that successfully describes it.  
Remarkably, the biologically relevant parameters to model transcription and translation fall into this regime; also, finite size effects are intrinsic to these biological processes, as genes and mRNAs are obviously far from the thermodynamic limit. Furthermore, we show that another variant of the exclusion process, the TASEP with dynamical defects (ddTASEP~\cite{turci_transport_2013, sahoo_asymmetric_2016, Waclaw2019}), used to model the impact of mRNA secondary structures on ribosome movement or DNA methylation on RNAP traffic, can be mapped onto the `pausing TASEP' (pTASEP), allowing us to solve both models with our refined framework.

\section{\label{sec:level1}The pausing TASEP (and TASEP with dynamical defects)}

\subsection{The Totally Asymmetric Simple Exclusion Process}
In the totally asymmetric simple exclusion process, particles stochastically hop forward with a hopping rate $\varepsilon$ on a discrete unidimensional lattice, subject to the condition that the next site is empty. 
The TASEP can be envisaged with two types of boundary conditions. 
With periodic boundary conditions, the particle density $\rho$ (ratio between the total number of particles $N$ and the number of sites of the lattice $L$) is fixed and the lattice can be represented as circular, see Fig.~\ref{fig:transcription_translation_model}(c). 
With open boundary conditions, the density is not fixed but fluctuates around a steady-state average value induced by the boundaries: particles can enter the lattice with a rate $\alpha$ and leave the lattice with a rate $\beta$. 
If all sites of the lattice have the same hopping rate $\varepsilon$, the TASEP is one of the rare non-equilibrium systems that can be solved exactly~\cite{Derrida1992, blythe2007nonequilibrium}. Mean-field approaches, which neglect correlations between neighbouring sites, provide a useful tool to approximate the dynamics of the TASEP.
In the thermodynamic limit $L\xrightarrow{}\infty$ the phase diagram predicted by the mean-field solution becomes exact~\cite{Derrida1992, blythe2007nonequilibrium}. 
When particle hopping is non-homogeneous and depends on the the particular site $i$, mean-field approaches can still provide reasonable predictions of particle current and density profiles~\cite{Shaw2003, erdmann2020key}, and an approximation based on the power series approximation of the master equation has been proposed recently~\cite{szavits2018deciphering, szavits2018power, ciandrini2023tasepy}.
The current in the mean-field approximation for the standard TASEP with particle density $\rho$ is equal to
\begin{equation}\label{eq:standard_TASEP}
    J_0 = \varepsilon \rho (1- \rho) \,,
\end{equation}
where we use the subscript $0$ to indicate the system with zero paused particles or dynamical defects, the meaning of which will be clear in the following sections. 
In open boundary conditions, the entrance and exit boundaries fix the steady-state density which is given by the following relations: 
\begin{equation}
\rho =
    \left\{
    \begin{aligned}
        \rho_{LD} &= \alpha /\varepsilon \quad  &\alpha < \beta, \alpha < \varepsilon/2\\
        \rho_{HD} &= 1 - \beta / \varepsilon \quad  &\beta < \alpha, \beta < \varepsilon/2\\
        \rho_{MC} &= 1 / 2 \quad  &\alpha, \beta \geq \varepsilon/2 \,.
    \end{aligned}
    \right.
    \label{bulk density}
\end{equation}
Each of these cases corresponds to a specific phase: `low density' (LD), where the density is set by the in-rate $\alpha$; `high density' (HD), governed by the out-rate $\beta$, and `maximum current' (MC) where the flow is limited by collisions in the bulk.
Implicitly, Eq.~(\ref{bulk density}) also defines what is known as the TASEP 'phase diagram', which attributes a phase to each point in the $(\alpha/\varepsilon,\beta/\varepsilon)$ parameter plane. More information on the TASEP can be found in many available reviews, for instance~\cite{blythe2007nonequilibrium, Chou2011}.

\subsection{The pausing TASEP (pTASEP)}
\begin{figure}[ht]        
    \includegraphics[width=\columnwidth]{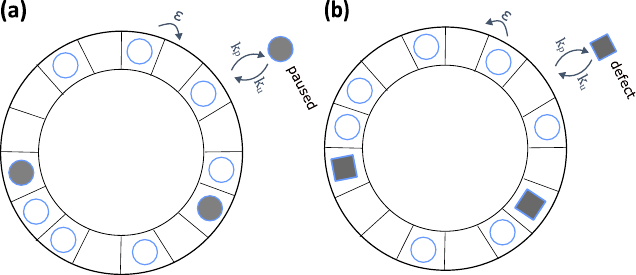}
    \caption{\label{fig:models} 
    Sketch of the TASEP with paused particles (pTASEP) in (a) and of the TASEP with dynamical defects (ddTASEP) in (b). 
    In the pTASEP, particles enter and leave a paused state with the rates $k_p$ and $k_u$, respectively. In the ddTASEP (b), defects bind and unbind to holes, impeding the traffic flow on the lattice. These systems are equivalent under a particle-hole mapping ($\rho \rightarrow 1-\rho$).}
\end{figure}

We investigate an exclusion process in which individual particles enter a paused state at a rate $k_p$, then exit this condition at rate $k_u$. 
As in the standard TASEP, the movement of particles occurs through hopping to the next unoccupied site, at rate $\varepsilon$. This variant of the TASEP has been introduced in~\cite{Klumpp2008} and then analytically investigated in~\cite{Wang2014}, where the authors developed a mean-field approximation. 
Our ultimate goal is to study the impact of antibiotics on ribosome elongation~\cite{Kavcic2020}. To this end we use the pTASEP, where pausing mimics the binding of antibiotics during translation elongation. The pausing/unpausing rates $k_p$ and $k_u$ thus reflect the rates at which antibiotics bind to the ribosome, and are hence expected to be related to the intracellular antibiotic concentration $[C]$ and its binding affinity $k_{on}$ as $k_p = k_{on} [C]$. Once in the paused state, particles can become active again at a rate $k_u$, simulating antibiotics unbinding from the ribosome. Figure~\ref{fig:models}(a) provides a visual summary of these dynamics. 
Unless otherwise stated, and without loss of generality, we fix the particle hopping rate to $\varepsilon = 1$ s$^{-1}$. This is equivalent to expressing the other rates in units of $\varepsilon$.

\subsection{Mapping between pausing particles and dynamical defects}
The TASEP with dynamical defects is another flavour of the TASEP, first introduced in~\cite{turci_transport_2013} for a single defect, and then extended in~\cite{sahoo_asymmetric_2016, Waclaw2019}. In the ddTASEP, defects bind and unbind to empty sites, obstructing the progression of particles. Such 'roadblocks' could model, for instance, the consequences of traffic lights in vehicular traffic, regulatory proteins in  transcription, or secondary structures in translation.

Even though the dynamics of pTASEP and ddTASEP is different, the two models are equivalent through a mapping between particles (occupied sites) in one model to the holes (unoccupied sites) in the other.
Indeed, in the same way that pTASEP consists of advancing particles which are paused/unpaused, the ddTASEP can be seen as receding holes which are blocked/unblocked, as shown in Fig.~\ref{fig:models}. Therefore the two models are equivalent (upon exchanging $\rho \rightarrow 1-\rho$).

In the remainder of the article we will focus on the pTASEP, but all results can be directly applied to the ddTASEP via this particle-hole symmetry.

\section{\label{sec:level1}Previous results and extension of the model }
\subsection{\label{sec:level2}Mean-field approach}
The pTASEP consists of two coupled dynamical processes. The first one is the standard exclusion process, reviewed in the previous section, the dynamics of which occurs far from equilibrium. The second process is the pausing and unpausing of particles which, instead, is at equilibrium. 
As a consequence, in steady state, the fraction of time a particle spends in the paused/unpaused states is directly given by the equilibrium relationships 
\begin{equation}\label{eq:fraction_paused_particle}
    f_p = \frac{k_p}{k_p + k_u} \,,
    \qquad
     f_a = \frac{k_u}{k_p + k_u} \,. 
\end{equation}
Thus, $f_p$ denotes the fraction of paused particles (i.e. paricles that cannot move due to their internal state), and $f_a$ the fraction of active (non-paused) particles, $f_p + f_a = 1$.
Wang {\it et. al}~\cite{Wang2014} developed a mean-field approximation for the pTASEP by assuming that only particles that are both active (not paused) and not blocked by a paused particle ahead, contribute to the current $J$. 
The mean-field solution of the current obtained in~\cite{Wang2014} reads:
\begin{equation}\label{eq:Wang} 
    J_\text{MF} = \varepsilon \rho f_\text{J}(1 - \rho),
\end{equation}
where 
\begin{equation*}
    f_\text{J} := \frac{k_u}{k_u + \hat{k}_p}\,,
    \qquad \hat{k}_p := k_p + \varepsilon \rho f_{p}\,.
\end{equation*}
This reflects the fact that a particle contributing to the current will now longer do so after becoming paused (rate $k_p$) or after moving (rate $\varepsilon$) to encounter a paused particle (probability $\rho f_p$). In the latter case the particle, although still active, becomes blocked from advancing further. 

These considerations enter the theory introduced in Wang {\it et. al}~\cite{Wang2014}, summarised in Eq.~(\ref{eq:Wang}).

In this formulation, $f_\text{J}$ can be interpreted as an effective probability of a particle contributing to the current, obtained from the typical time a particle spends in the paused state $(1/k_u)$ and from the effective time $(1/\hat{k}_p)$ spent in its current-contributing state. 

\begin{figure}[b]
    \includegraphics[width=\columnwidth]{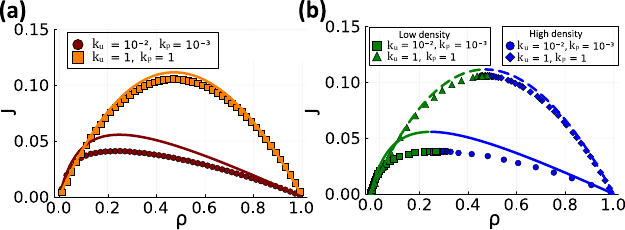}
    \caption{Comparison between stochastic simulations and mean-field approximation in (a) periodic boundary conditions and (b) open boundary conditions. When $k_u$ and $k_p$ are approximately equal to $\varepsilon$, shown in orange squares in (a) and green triangles for low density and blue diamonds for high density in (b), the agreement between simulations and theory is good. Deviations start to appear when the timescales of hopping and pausing kinetics differ significantly, as shown in dark red circles in (a) and green squares and blue circles in (b). In both plots $k_p$ and $k_u$ are expressed in units of $\varepsilon$ and $L = 1000$.}
    \label{Fig:periodic_and_open_boundaries}
\end{figure}

Figure~\ref{Fig:periodic_and_open_boundaries}(a) compares the prediction of this mean-field approach and the outcome of stochastic simulations (see Appendix~\ref{app:sims}) with periodic boundary conditions, for two different choices of the parameters. 
Equation~(\ref{eq:Wang}) clearly captures the overall behaviour seen in stochastic simulations. However, while remaining indicative of the numerical results, the analytical solution becomes less accurate when $k_u$ and $k_p$ decrease. 
In Section~\ref{sec:finite} we will investigate more thoroughly the accuracy of mean-field predictions for a broad range of parameters.

\subsection{\label{sec:level2}Expanding the mean-field approach to open boundary conditions}

Since the fundamental relation $J(\rho)$ is known for the pTASEP, see Eq.~(\ref{eq:Wang}), it is possible to extend the mean-field solution to the open boundary condition, which reveals how the rates $\alpha$ and $\beta$ drive the different phases. 
In order to do so we exploit the maximum current principle for boundary-induced phase transitions, developed by Krug~\cite{Krug1991}. It states that
\begin{equation}
    J = \max_{\rho \in [\rho_{L+1}, \rho_0]} J(\rho) 
    \, .
\end{equation}
where $0$ and ${L+1}$ refers to two (fictitious) reservoir sites at either end of the segment by which the lattice is extended. Their densities are chosen in order to emulate the desired entry and exit rates, respectively, assuming furthermore that the dynamics between the reservoirs and the lattice follows the same microscopical rules as in the bulk. 

Therefore the current from the reservoir with density $\rho_0$ and the first site is $\varepsilon \, \rho_0 \, \,  f_\text{J} (1-\rho_{1})$, which is to be assimilated to the current set by the desired entry rate, $\alpha (1-\rho_1)$. Thus $\alpha = \varepsilon \, \rho_0 \,  f_\text{J}$. Similarly, the exit current is fixed by the fraction of active particles that can leave the last site with rate $\beta$, that is $\beta \rho_L f_a$. Of note, all active particles on the last site contribute to the current as they cannot be blocked by traffic and paused particles downstream. This exit current is to be emulated by the density $\rho_{L+1}$ of the right reservoir. Since the latter implies a current $ \, \varepsilon \, \rho_L \, f_\text{J} \, (1-\rho_{L+1})$, this thus requires $\beta = \varepsilon (1-\rho_{L+1}) f_\text{J}/f_a$.
Consequently, the reservoir densities $\rho_0$ and $\rho_{L+1}$ are directly set by the entry/exit rates, respectively. 

\begin{figure}[t]
    \includegraphics[width=\columnwidth]{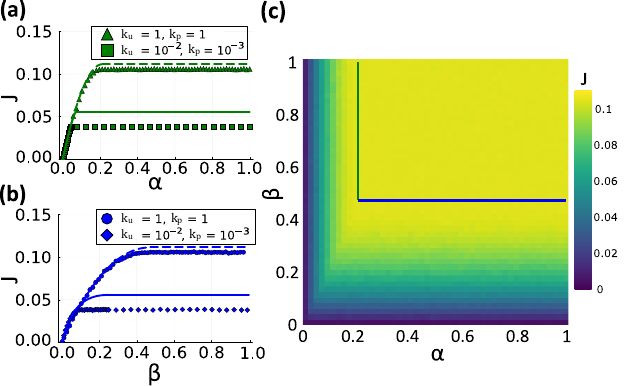}
    \caption{Critical values and phase diagrams. Panels (a) and (b) show the current as a function of $\alpha$ and $\beta$ respectively for $L=1000$. Similar to Fig.~\ref{Fig:periodic_and_open_boundaries}, the pTASEP captures the behavior, when the elongation rate is of the order of the pausing and unpausing rates, here represented by the dashed line. On the contrary, the prediction, depicted by the solid line, becomes worse for lower values of $k_u$ and $k_p$ (expressed in units of $\varepsilon$) compared to the elongation rate. Panel (c) depicts the entire phase diagram for $k_u = k_p = \varepsilon = 1$. Straight lines correspond to the predictions of $\alpha_{crit}$ and $\beta_{crit}$ while the heatmap represents the outcome of the simulated current.}
    \label{fig:critical_values}
\end{figure}

The phase diagram is obtained from the maximal current principle, which is implemented setting the derivative in Eq.(~\ref{eq:Wang}) to zero:
\begin{align}
    J &= J_{LD} \quad \text{for} \quad \rho_0 < 1 - \rho_{L+1}, \:\rho_0 < \rho_{max} \\
    J &= J_{HD} \quad \text{for} \quad \rho_0 > 1 - \rho_{L+1}, \:\rho_{L+1} > \rho_{max} \\
    J &= J_{MC} \quad \text{for} \quad \rho_0 \geq \rho_{max}, \: \rho_{L+1} \leq \rho_{max}
\end{align}
with 
\begin{equation}\label{eq:rho_max}
    \rho_{max} := - \chi + \sqrt{\chi^2+\chi} \hspace{4mm} \text{and} \hspace{4mm} \chi := \frac{k_p + k_u}{\varepsilon f_p}.
\end{equation}
In Fig.~\ref{Fig:periodic_and_open_boundaries}(b) we show the fundamental diagram $J(\rho)$ for the open boundary conditions, where the density $\rho$ is varied by tuning $\alpha$ and $\beta$. The left branch (LD) is obtained by varying $\alpha$ at a fixed $\beta$ and, conversely, the right branch (HD) by fixing $\alpha$ and changing $\beta$. The two branches meet in the maximal current phase, at $\rho=\rho_{max}$, i.e. where the current is limited by the bulk of the system and no longer by the boundaries. The transition from low density to maximal current occurs when $\rho_{0} = \rho_{max}$, leading to $\alpha_{crit} = f_\text{J} \varepsilon \rho_{max}$. Similarly, the transition between high density and maximum current takes place at $\rho_{L+1} = \rho_{max}$ which leads to $\beta_{crit} = \varepsilon (1-\rho_{max}) f_\text{J}/f_a $. In Fig.~\ref{fig:critical_values} (a)-(b), we compare $J(\alpha)$ and $J(\beta)$ obtained following this method, for different values of $k_u$ and $k_p$. 
Similar to the periodic boundary condition case, the analytical predictions become worse for small values of $k_u/\varepsilon$ and $k_p/\varepsilon$. 
Figure~\ref{fig:critical_values}(c) displays the phase diagram for $k_u/\varepsilon =  k_p/\varepsilon = 1$, the heatmap representing the outcome of simulations. Although, the fundamental diagram in Fig.~\ref{Fig:periodic_and_open_boundaries}(b), depicted as green squares and blue diamonds, show a re-scaled TASEP, the phase diagram  in Fig~\ref{fig:critical_values}(c) is shifted. This is because the pausing and unpausing dynamics act differently on the initiation and termination site which can also be seen in the aforementioned equations for $\alpha$ and $\beta$. 

In the next section we will show that the mean-field predictions, provided in~\cite{Wang2014}, capture the fundamental relation $J(\rho)$ in most of the parameter space at least qualitatively. However, the mean-field prediction fails in biologically relevant regimes for transcription and translation due to severe finite size effects. In the remainder of the article, we propose a correction for both the periodic and open boundary case.

\section{\label{sec:finite}Enhanced description of finite-size effects in transcription and translation}
In this section, we introduce the idea that the pTASEP can be used to mimic the effect on ribosome dynamics of a class of antibiotics interfering with the advancement of the ribosomes during elongation. We find that, in the parameter regime of interest, the mean-field approach fails, and we show that this is due to severe finite-size effects. We then develop an improved theory for this regime, both for periodic and open boundary conditions.
\subsection{\label{sec:antibio}The transcription and translation parameter space}

We ballpark the range of parameter values and emphasise that relevant parameters of transcription and (antibiotic-inhibited) translation fall in regimes that are not necessarily well described by previous approaches.\\
{\noindent \it Transcription.} Single-molecule experiments on transcription in vitro have revealed that RNAPs enter and exit a paused state with rates that are between 0.07 - 0.15 s$^{-1}$ and approximately 1 s$^{-1}$, respectively~\cite{Klumpp2008}. The elongation rate of RNAPs ranges from 20 to 80 nucleotides per second, resulting in a timescale separation between pausing and unpausing dynamics of one to three orders of magnitude~\cite{vogel1994, vogel1995, epshtein2003}. Moreover, the ddTASEP has been used in the literature to describe the impact of DNA blockages on RNAP traffic. These dynamical blockages can be any regulatory proteins binding and unbinding the DNA template, ranging from histones~\cite{vandenberg_crowding-induced_2017} to methylated-DNA binding protein~\cite{cholewa-waclaw_quantitative_2019}. For instance, the MeCP2 binding protein has a $k_p$ and $k_u \approx 0.04/s$, to be compared with the RNAP elongation rate mentioned above, i.e. $\varepsilon \sim 20-80$ nucleotides/s.\\
{\noindent \it Translation.}
There exist different classes of antibiotics affecting cell growth by a plethora of disparate mechanisms of action.
Here we are interested in antibiotics hindering protein synthesis, and in particular the ones affecting ribosome dynamics during the elongation stage~\cite{Kavcic2020, wilson_ribosome-targeting_2014}.
Chloramphenicol, erythromycin and tetracycline are examples of antibiotics acting in such a way. 
The binding and unbinding rates of these translation inhibiting antibiotics
are several orders of magnitude lower than the elongation rate of the ribosome~\cite{Greulich2015, Dai2016, tritton_ribosome-tetracycline_1977, wilson_ribosome-targeting_2014}. 

In the translation process, the elongation rate is between 10-20 amino acids per second~\cite{Dai2016}. The binding and unbinding times of the aforementioned antibiotics are several orders of magnitude lower then the elongation rate, thus leading to a timescale separation between the translation and antibiotics dynamics. For instance, chloramphenicol has a binding rate constant of ($k_{on} = 5.6 \times 10^{-4}  \mu M^{-1} s^{-1}$) and an unbinding rate of ($k_u =1.23 \times 10^{-3} s^{-1}$)~\cite{harvey1980}. As typical sublethal doses of antibiotics range from $2$ $\mu M$ to $12$ $\mu M$~\cite{Scott2010}, the timescales are up to five orders of magnitude apart.\\

Thus, in order to investigate transcription and translation in a biologically relevant regime, we need to study the regime $k_u, k_p \ll \varepsilon$.

\subsection{\label{sec:PBC}Periodic boundary conditions}

Mimicking the aforementioned effect of antibiotics, we have simulated the pTASEP in periodic boundary conditions by varying $k_p$ over several orders of magnitude. 

\begin{figure}[ht]
    \includegraphics[width = \columnwidth]{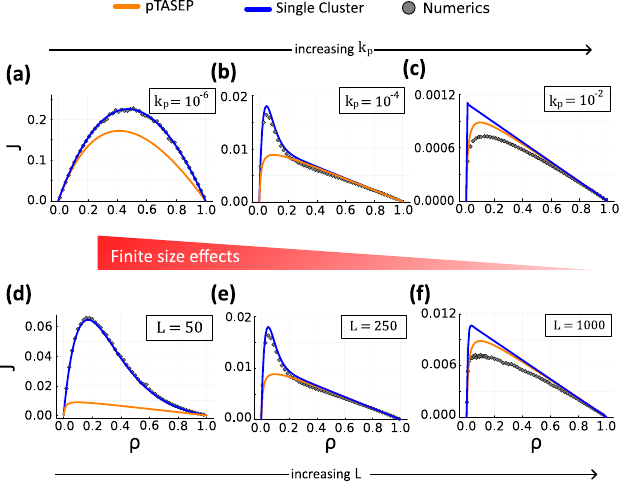}
    \caption{\label{fig:PBC_sim} Relevance of the finite size effects in the $J(\rho)$ relation. The pTASEP and single cluster solution are shown in blue and orange respectively, whereas the numerical simulation is in grey. The figures (a) to (c) show the emergence of finite size effects by varying $k_p$ from $10^{-6}$, $10^{-4}$ and $10^{-2}$, respectively, while keeping $L=250$ and $k_u=10^{-3}$. The figures (d) to (f), show the same effect emerging by changing the system size $L$, while keeping $k_u = 10^{-3}$ and $k_p = 10^{-4}$ fixed.
    }
\end{figure}

By comparing the model to the outcome of simulations we notice that the mean-field prediction fails, not only quantitatively but also qualitatively, when decreasing $k_p$, see Fig.~\ref{fig:PBC_sim}(a)-(c). 
We observe that the discrepancy between the mean-field prediction given by Eq.~(\ref{eq:J_PBC}) and simulations arises when there is likely no paused particle on any of the finite number of sites. The expected number of paused particles is given by $N_p := f_p \rho L$, which depends on the rates $k_p$ and $k_u$, on the particle density and the lattice length.
The impact of the blockages imposed by paused particles, and thus Eq.~(\ref{eq:J_PBC}), is relevant only when $N_p$ is at least comparable to $1$, which is when the total number of particles is sufficiently large ($\rho L \geq f_p^{-1}$).
Otherwise, we expect finite sizes to be relevant.

In Fig.~\ref{fig:PBC_sim} we show $J(\rho)$ by varying the microscopic pausing rate $k_p$, panels (a)-(c), or by changing the lattice size, panels (d)-(f). We notice that, when the number of expected paused particles $N_p$ is order $1$ or smaller the current exceeds the pTASEP mean-field predictions (compare the simulation points and the orange curve). For instance, at the peaks of panels (b) and (e), $N_p \sim 1$.  With paused particles being scarce, Fig.~\ref{fig:PBC_sim}(a)-(b)-(d)-(e), the system behaves as a standard TASEP (i.e. without paused particles) for extensive amounts of time, with a current therefore exceeding the mean-field predictions. For larger densities, the probability of observing paused particles increases ($N_p > 1$) and the mean-field argument captures the outcome of the simulations. 

Understanding the role of the system size is therefore crucial: finite-size effects in the pTASEP are intrinsically related to the pausing-unpausing dynamics via $N_p$, and are significant in biologically relevant regimes. %
To investigate finite-size effects, we juxtapose Fig.~\ref{fig:PBC_sim}(d)-(f) by using the same dynamical parameters as in the central panel (b), but with progressively larger lattice sizes (up to $L=1000$). We observe that the `bump' in the current disappears for large lattices, whereas it is amplified in small lattices. Additionally, deviations from the mean-ield predictions become more pronounced in smaller lattices.
Thus, simulations clearly show that there is a size dependent behavior, and that mean-field approaches greatly underestimate the current when finite size is relevant (i.e. when $N_p \leq 1$). We address this situation by formulating a two-term expression: one term accounts for the current produced in the absence of paused particles (the unpaused state), weighted by the probability \(P_0\), and the other term captures the current in the presence of at least one paused particle (the paused state), weighted by the complementary probability \(1 - P_0\). Since the particle number $\rho L$ is constant, as well as the average ratio of active particles $f_a$ (at steady state), $P_0$ is given by
\begin{equation}
    P_0= (f_a) ^{\rho L} = \left(\frac{k_u}{k_u+k_p}\right)^{\rho L}
\end{equation}

and the current equation reads

\begin{equation}
    J=P_0 J_0 + (1-P_0) J_{\text{p}} \,,
    \label{eq:J_PBC}
\end{equation}

where \(J_0\) denotes the current in the unpaused state (no paused particle on the lattice), and \(J_{\text{p}}\) corresponds to the current when at least one particle is in the paused state. As we will see, in the regime of rare and long pauses, a single pausing event typically leads to the formation of large clusters.

In the following, we will first develop a single-cluster approximation and obtain a solution for $J_p$ and $J_0$ to better understand the regime dominated by finite size effects, and compare this to results from the literature. Eventually we provide a more phenomenological solution extending our approach to all regimes.

\subsubsection{Single-cluster approximation for $J_\text{p}$ and $J_0$}
\label{sec:single-cluster}
\begin{figure}
    \centering
    \includegraphics[width = \columnwidth]{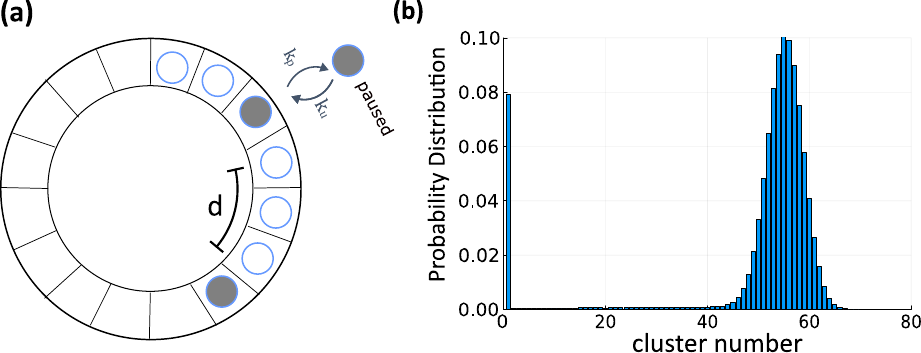}
    \caption{Justification for the single cluster approximation: 
    (a) illustrates the single cluster regime, where distance \(d\) is the part of the cluster that breaks off. When \(\varepsilon \gg k_u, k_p\), the break off travels around the circle before a new unpausing event. At low densities, a single paused particle can block the cluster, making \(d\) cover the entire cluster, which upon an unpausing event, leads to the system with no paused particles. Thus, the system behaves like a standard TASEP.
    (b) shows the probability distribution of the number of clusters for \(L = 250\), \(k_u = 10^{-3}\), \(k_p = 10^{-6}\), \(\rho = 0.33\). The distribution is bimodal: the right side is Gaussian, confirming the standard TASEP phase (unpaused state), while the left side shows a large peak indicating a single cluster of particles.}
    \label{fig:single_cluster_justification}
\end{figure}

When the particle movement is much faster than the pausing dynamics (i.e. in the biologically relevant regime $k_u,k_p \ll \varepsilon$) it is reasonable to first assume that all particles cluster together behind a particle in the paused state. In this state, unpausing events of the leading particle determine an intermittent current $J_p$ established by the detaching fronts. 
When finite-size effects are relevant (the number of paused particle, $N_p$ is order $1$), there is a non-negligible probability of dissolving the complete cluster, which effectively goes to zero for high density. In this case, the system will relax from the paused state to the unpaused state, with a particle current $J_0$.
In the single-cluster steady state, paused particles are separated by $d$ active particles as shown in Fig.~\ref{fig:single_cluster_justification}(a). When the leading particle becomes unpaused, $d+1$ particles detach from the cluster, representing an effective density $(d+1)/L$. Due to the aforementioned timescale separation between the two dynamics and the modest lattice lengths of the systems of our interest, we make the hypothesis that these particles travel the $L(1-\rho)$ empty sites without becoming paused and rejoin the cluster from the other end. This sub-cluster detachment occurs intermittently, with a typical time between events given by $1/k_u$, the inverse of the unpausing rate of the particle at the front of the cluster.
Putting all this together, in the paused state the mean particle current reads 
\begin{equation}
    \label{eq:Jsingle_cluster}
    J_{p}=k_u(1-\rho)(d+1) \,.
\end{equation}

The number of particles $d$ detaching together with the leading particle can be computed explicitly as the expected value of the number of negative results (finding a free particle) of a Bernoulli experiment before a successful one (finding a paused particle). If $x$ is said random variable then we know that it is distributed as a negative binomial distribution and its expected value is:

\begin{equation*}
    \left<x\right>=\sum_{i=0}^{+\infty}(f_a)^i(1-f_a)^i \,i \,.
\end{equation*}
However, as the cluster is finite, we constrain $d+1 \leq N$, i.e. $d \leq \rho L - 1$. 

We thus obtain

\begin{equation*}
    d=\sum_{i=0}^{\rho L-2}(f_a)^{i}(1-f_a) i +\sum_{i=\rho L-1}^{+\infty}(f_a)^{i}(1-f_a)(\rho L-1) \,,
\end{equation*}

where the last sum bounds $x$ with $(\rho L-1)$ in the Bernoulli experiment. After using the geometric series (and semi-series) identities we obtain:

\begin{equation}
    d=\frac{k_u}{k_p}\left(1-\left(\frac{k_u}{k_u+k_p}\right)^{\rho L}\right) \,,
\end{equation}

which reduces to $k_u/k_p$ for clusters of infinite length. The value of $d$ can then be plugged into Eq.~(\ref{eq:Jsingle_cluster}) to evaluate~$J_p$.\\

Instead of simply considering the steady-state TASEP current $J_0$ given in Eq.~(\ref{eq:J_PBC}), here we first model the dissolution of the cluster and study how the density profiles relax to the homogeneous profile of the standard TASEP. Put differently, we  compute the time-dependent particle current $J_0(t)$ which relaxes to the steady state value given by Eq.~(\ref{eq:J_PBC}).
The evolution of density profiles of the exclusion process can be studied in the hydrodynamic limit~\cite{blythe2007nonequilibrium}, considering $\rho(x,t)$ as a continuous variable in space and time. It has been shown that $\rho(x,t)$ is a solution of the generalised Burgers' equation. 
By this approach we obtain a time dependent density profile $\rho(x,t)$ (see Appendix~\ref{app:current}), which in turn gives us the instantaneous current \begin{equation}
j(t)=\frac{\varepsilon}{L}\int_{0}^{L}\rho(x,t)\left(1-\rho(x,t)\right)\,dx \,,
\label{current_integral}
\end{equation}
which is the space-averaged current at time $t$.

We are interested in the current from time $0$, that is when the fully packed cluster (density $1$) starts to dissolve due to the absence of paused particles, to the expected time required for a new pause, which on average will occur after a time $\tau=\frac{1}{\rho L k_p}$ that corresponds to the inverse of the rate of pausing any of the $\rho L$ particles.  
By solving the Burgers' equation with the method of characteristics, we obtain $\rho(x,t)$ and we compute the integral Eq.~(\ref{current_integral}). We obtain:

\begin{equation}
    j(t)=
    \begin{cases}
        \frac{\varepsilon^2}{6L}t & \text{if} \ t \in [0,\frac{\rho L}{\varepsilon})\\
        \varepsilon \rho \left(1-\frac{2}{3}\sqrt{\frac{\rho L}{\varepsilon t}} \right) &
        \text{if} \ t\in \left[\frac{\rho L}{\varepsilon},\frac{L}{4\rho\varepsilon}\right)\\
        \varepsilon\left[\rho(1-\rho) -\frac{L}{48\varepsilon^2 t^2}\right] & \text{if} \ t\in \left[\frac{L}{4\rho\varepsilon},+\infty \right) \,.
    \end{cases}
\end{equation}

The full derivation of the instantaneous current $j(t)$, along with an explicit formula for the density profile at all times is given in Appendix~\ref{app:current}. 
Thus, $J_{0}$ in Eq.~(\ref{eq:J_PBC}) is given by the function
\begin{equation}
    J_0(\tau)=\frac{1}{\tau}\int_{0}^{\tau}j(t)\,dt
    \label{eq:J0}
\end{equation}
evaluated at $\tau=\frac{1}{\rho L k_p}$.\\

We now have all components of Eq.~(\ref{eq:J_PBC}), which can thus be tested against numerical results.
In Fig.~\ref{fig:PBC_sim} we plot the result of Eq.~(\ref{eq:J_PBC}), using $J_0$ from Eq.~(\ref{eq:J0}) and $J_p$ from the single-cluster approximation Eq.~(\ref{eq:Jsingle_cluster}), compared to both simulations and the mean-field theory. Our framework not only correctly describes the regimes when finite size effects are relevant but also captures them quantitatively.

\subsubsection{Comparison with previous approaches}
We now further investigate and compare the previously developed mean-field arguments for the ddTASEP and the pTASEP from~\cite{Wang2014,Waclaw2019} together with the single-cluster approximation introduced in this work. 

\begin{figure}[!b]
    \includegraphics[width=\columnwidth]{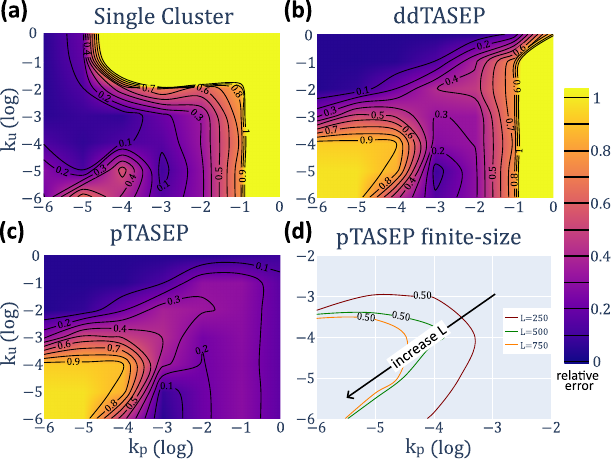}
    \caption{\label{fig:heatmaps} Relative error between the maximal value of the simulated current and the predicted value at the corresponding particle density, computed as $|J - J_{\text{sim}}| / |J_{\text{sim}}|$, where $J$ represents the analytical solution for the current and $J_{\text{sim}}$ is the simulated current. Panel (a) shows a heatmap for the error in the single-cluster solution given by Eq.(~\ref{eq:Jsingle_cluster}). Panels (b)-(c) show the ddTASEP and pTASEP using the mean-field approach, respectively. In panel (d) we emphasise how the parameter region where the approach fails shrinks with increasing the system size. We notice how previous approaches cannot capture the regime of long ($k_u \ll \varepsilon$) and rare ($k_p \ll \varepsilon$) pauses. For panel (a)-(c), $L=250$.
}
\end{figure}

Our single-cluster approximation succeeds in predicting the system's behaviour in cases of strong finite size effects, as shown in Fig.~\ref{fig:PBC_sim}. We now systematically asses the quality of the single-cluster approximation in the $\{k_u, k_p \}$ parameter space by comparing the particle current predicted by our approach to that obtained from simulations. Figure~\ref{fig:heatmaps}(a) shows a `heatmap' of the relative error between $J_{max}$ obtained with simulations and with Eq.~(\ref{eq:J_PBC}). 
For comparison we plot the same quantity for previously proposed mean-field approaches: the ddTASEP as studied in~\cite{Waclaw2019} -Fig.~\ref{fig:heatmaps}(b)- and $J_\text{MF}$~\cite{Wang2014} for the pTASEP -Fig.~\ref{fig:heatmaps}(c). 
The models from the literature fail to capture finite-size effects occurring in the bottom left region, compared to the single cluster model introduced in the previous section. The abruptly increasing error for large $k_u$ and $k_p$, for the single-cluster model, is due to the failure of the single-cluster assumption in regimes where the timescale separation between those rates and $\varepsilon$ is less clear. In this regime, the single-cluster approximation obviously fails, being in a sense complementary to previous mean-field descriptions of the pTASEP and ddTASEP that cannot capture finite-size effects.

To further emphasise the fact that former approaches break down because of finite-size effects, in Fig.~\ref{fig:heatmaps}(d) we show the contour lines at $50\%$ relative error of $J_\text{MF}$ for different values of $L$, pointing out how the region with the largest disagreement (small $k_u$, $k_p$ with respect to $\varepsilon$) shrinks as the system size increases. 

\subsubsection{Extending the mean field in a multi-cluster regime}
\label{sec:multi-cluster}
The single cluster approximation fails for increasing $k_u$, $k_p$, and $L$. On the one hand, when $k_u$ increases any cluster will break up before it can collect all the particles, leading to the emergence of multiple clusters. On the other hand, an increase in $k_p$ results in more frequent pauses, resulting in the inability for all the particles to coagulate into a single cluster during the time $1/k_u$. The system size ($L$) further exacerbates these issues by extending the time required for all the particles to form a single cluster. Consequently, both terms in Eq~(\ref{eq:Jsingle_cluster}) fail due to the emergence of multiple clusters.

\begin{figure}[!b]
\includegraphics[width=0.9\columnwidth]{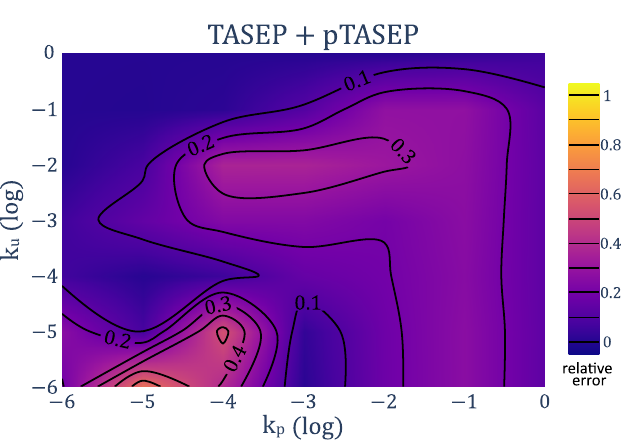}
\caption{\label{fig:TASEP_pTASEP} Relative error (at the maximum) between simulations and theory, Eq.~(\ref{eq:pTASEP_TASEP}). The extended mean-field equation works reasonably well for a large range of $k_u$ and $k_p$. In this plot $L=250$.}
\end{figure}

To address these limitations, we propose an effective mean-field approach. Starting from Eq.~(\ref{eq:Jsingle_cluster}), we replace the single-block current $J_p$ with the mean-field current $J_\text{MF}$ given by Eq.~(\ref{eq:Wang}) and the current term with no paused particles $J_0$ with the standard TASEP current given by Eq. (\ref{eq:standard_TASEP}). The resulting expression for the current $J$ is:

\begin{equation}\label{eq:pTASEP_TASEP}
    J = P_0 J_{0} + (1-P_0) J_\text{MF}
\end{equation}

It is important to note that, while this equation does not explicitly depend on the system size, the weight $P_0$ does. As seen in Figure~\ref{fig:TASEP_pTASEP}, Eq.~(\ref{eq:pTASEP_TASEP}) works well over various orders of magnitude for $k_u$ and $k_p$, indicating the robustness of our approach.

\subsection{\label{sec:OBC}Open boundary conditions}
Treating open boundaries is considerably more complex than the periodic case, as the number of particles on the lattice is no longer constant over time. As a consequence, we can neither use the number of particles to compute probabilities nor assume that the pausing dynamics has reached its steady state.
Within the parameter range described in Section \ref{sec:antibio} one would expect the single-blockage phenomenology explored from the periodic boundary case translates to the open boundary case. The lattice can then be found in either of two states: a paused state (bearing similarities with the paused state in periodic boundary conditions), where the lattice is completely filled up to the site of the first paused particle, and a unpaused state, where the system behaves as a standard TASEP. As in the periodic case, the cluster in the paused state consists of paused particles interspersed with $d$ active ones. When the leading paused particle unpauses, it triggers the detachment of $d+1$ particles. This process repeats until the cluster dissolves, after which new particles can re-initiate on the lattice. 

Because of the time separation between $\alpha$, $k_u$, $k_p$ and $\varepsilon$, we can assume that particles quasi-instantaneously fill the lattice up to the position of the first particle becoming paused. Similarly, and for the same reason, particles downstream from the rightmost paused particle would quasi-instantaneously exit the lattice. 
Reproducing the reasoning followed to obtain Eq.~(\ref{eq:J_PBC}), we therefore introduce a two term approach: 
\begin{equation}
    J=\frac{\tau_0}{\tau_p+\tau_0}J_{0}+\frac{\tau_p}{\tau_p+\tau_0}J_{p} \,,
\label{eq:OBC_current}   
\end{equation}
where $\tau_0$ and $\tau_p$ are the average duration of the unpaused state and of the paused state, respectively. The unpaused state behaves as a standard TASEP. Thus, $J_0$ is given by the standard TASEP current Eq.(\ref{eq:standard_TASEP}).

We denote by $x$ the position at which, starting from the unpaused state, a particle transitions to the paused state. The position $x$ (in units of lattice sites) also corresponds to the total number of particles constituting the initial cluster. As a first approximation, we neglect the transient and assume the system enters steady state immediately after exiting the paused state. This leads to a homogeneous particle density, and since all particles have equal probability of pausing, the average position of the first paused particle is $\left< x \right> = L/2$.

Thus the total current of the paused state can be computed as the number of particles in the initial cluster divided by the typical time $\tau_p$ of its duration, $J_p = L/(2\tau_p)$. Equation~(\ref{eq:OBC_current}) then becomes:

\begin{equation*}
    J=\frac{\tau_0}{\tau_0+\tau_p}J_0+\frac{1}{\tau_0+\tau_p}\frac{L}{2}.
\end{equation*}

The first timescale $\tau_0$ can be computed by considering that the rate of switching from the unpaused state is proportional to the total number of particles times $k_p$, that is

\begin{equation*}
    \tau_0=\frac{1}{\rho L k_p}=\frac{\varepsilon}{\alpha L k_p}\,,
\end{equation*}

where in the last equivalence we consider a lattice in the LD phase ($\rho = \alpha/\varepsilon$).
We then compute the mean lifespan $\tau_p$ of the paused state following a numerical approach. This timescale can be smaller or comparable to the time necessary to reach an equilibrium distribution of the paused particle fraction $f_p$ given in Eq.~(\ref{eq:fraction_paused_particle}), and for this reason writing a closed form for $\tau_p$ is cumbersome. Thus, after the first pausing event leading to the formation of the extensive cluster, other particles can become paused, slowing down the restoration of the unpaused.  Eventually, the paused state will disappear when all particles will successively re-become active and exit the lattice together with the particles that were blocked behind them. The silent assumption behind our reasoning is that active particles which are not trapped in the cluster will exit the lattice without becoming paused again.

To develop our numerical method we first compute the time dependence of $f_p$ starting from a situation in which all particles in the cluster are active:

\begin{equation}
    f_p(t)=\frac{k_p}{k_p+k_u}\left(1-e^{-(k_p+k_u)t}\right)\,,
    \label{time_dip_prob}
\end{equation}

which asymptotically reaches the steady-state probability $f_p = \frac{k_p}{k_p+k_u}$. At a given time $t$, the mean number $d_t$ of particles `trapped' between the first and the second rightmost paused particles can be computed as the mean number of failures in a sequence of independent and identically distributed Bernoulli trials before another particle is found paused (success). This is computed as the average of a negative binomial distribution with rate of success $f_p(t)$ (Eq.~(\ref{time_dip_prob}), giving 
\begin{equation}
    d_t = \frac{1-f_p(t)}{f_p(t)}\,.
    \label{eq:d_t}
\end{equation}

The average lifetime $\tau(x)$ of a paused particle situated at $x$ is $t_- := 1/k_u$, and the expected number of particles between two paused particles is given by Eq.~(\ref{eq:d_t}) computed at $t = t_-$. We now have all the elements to numerically compute the typical time to free all particles blocked in the paused state. The number of particles freed after the leading particle becomes active, given after time $t_1 := 1\times t_-$, is given by $d_1 + 1$, where $d_1$ is the number of particles in between the first and the second paused particles. If $d_1+1$ is greater than $x$, we stop and estimate $\tau(x) = t_-$. Otherwise, we compute the next expected distances $d_i$ evaluating Eq.(\ref{eq:d_t}) at $t_i = i \times t_-$ until $\sum_{i=1}^j (d_i+1)>x$. Then, we approximate $\tau(x)$ as $j \times t_-$. 
Averaging $\tau(x)$ over all possible cluster lengths finally gives us the $\tau_p$ to be used in Eq.~(\ref{eq:OBC_current}).

\begin{figure}[!t]
\includegraphics[width=\columnwidth]{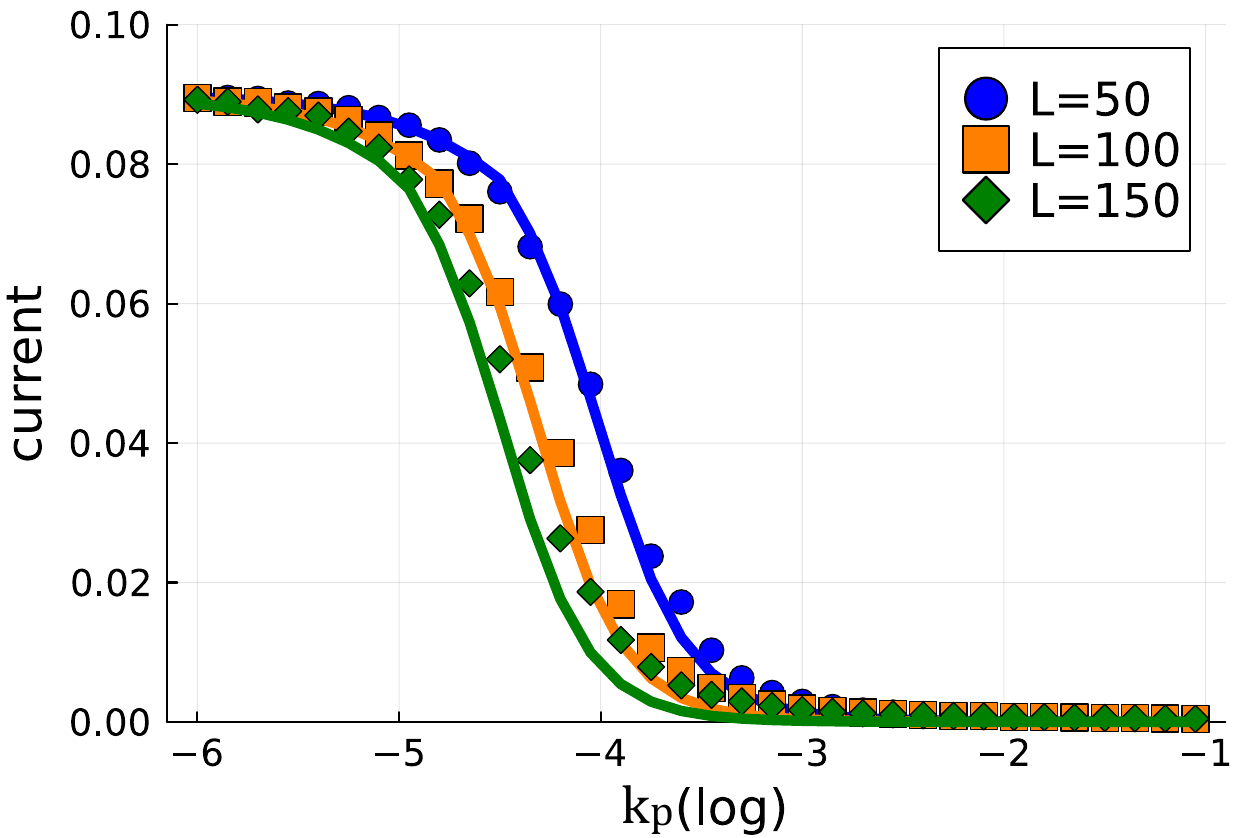}
\caption{\label{fig:plots_OBC}
Compares the current as a function of $k_p$ for different system sizes. Here $k_u = 10^{-3}, \alpha = 0.1,$ and $\beta = 1$. The solution given by Eq.~\ref{eq:OBC_current} works well for small system sizes and becomes worse for larger lattices.
}
\end{figure}

Our theory is in good agreement with the simulations, particularly for small lattices. In particular, in Fig.~\ref{fig:plots_OBC} we show how our two-state approximation Eq.~(\ref{eq:OBC_current}) correctly predicts the particle current $J$ even for small systems ($L=50$) and when the pausing rate $k_p$ is varied over several orders of magnitudes. 

\section{Discussions}
Motivated by the modes of action of several translation-inhibiting antibiotics, in this article we revise previous mean-field approaches modelling the impact of stochastically pausing particles in an exclusion process. To our knowledge, this variant of the totally asymmetric simple exclusion process, named pTASEP, has first been introduced to describe RNAP dynamics during transcription~\cite{Klumpp2008}, where the authors simulated the system in the regime of interest. A few years later, a mean-field approach has been introduced~\cite{Wang2014}, restricted to the periodic boundary case. 

We remark that this model is equivalent to another variant of the TASEP, the exclusion process with dynamical defects (ddTASEP): this model has been used in the past to provide a simplified description of the effects of secondary structures on mRNA translation~\cite{turci_transport_2013} and on the occlusion of RNAPs by DNA-binding proteins or nucleosomes~\cite{cholewa-waclaw_quantitative_2019, Waclaw2019, vandenberg_crowding-induced_2017}. The results shown in this article therefore hold for both the pTASEP and ddTASEP. By characterising mathematically the impact of antibiotics on ribosome dynamics, we effectively investigate a larger class of problems related to transcription and translation.  

After summarising the pTASEP mean-field theory available in the literature, Eq.~(\ref{eq:Wang}), we extend it to the open boundary case by means of the maximal current principle~\cite{Krug1991}. Thanks to this we obtain the critical values of entry and exit rates $\alpha$ and $\beta$, and establish the corresponding phase diagram (see Fig.~\ref{fig:critical_values}).

Intriguingly, in both periodic and open boundary conditions (see Figs.~\ref{Fig:periodic_and_open_boundaries}-\ref{fig:critical_values}), this mean-field approximation fails when $k_p/\varepsilon$ and $k_u/\varepsilon \ll 1$. We provide a few estimates of rates involved in transcription and translation, and point out that biologically relevant regimes are close to, or fall into, this region of poor agreement between theory and simulations. Notably, our model treats particles as occupying a single site, whereas RNAPs and ribosomes each span multiple sites. Accounting for their true footprint would reduce the maximum particle load on the lattice, amplifying finite-size effects and driving both transcription and translation further into the finite-size regime~\cite{keisers2025pausedtranslationmodeltranscript}. Corroborated by a systematic exploration of the parameter space (Figs.~\ref{fig:PBC_sim}-\ref{fig:heatmaps}), we conclude that severe finite-size effects cause the inadequacy of existing mean-field approaches when $k_p/\varepsilon$ and $k_u/\varepsilon \ll 1$. 
In contrast to the standard TASEP, where the mean-field theory performs well even with relatively small lattices (of the order of $100$ sites), and where the agreement with simulations is independent of hopping and boundary rates, the pTASEP exhibits finite-size effects that depend on parameter values. While the thermodynamic limit remains a reasonable approximation for finite systems, the required system size to achieve this condition is closely tied to the separation of timescales between the particle number (i.e the system size), pausing/unpausing dynamics and particle hopping. 

We develop a single-cluster approximation (Section~\ref{sec:single-cluster}) to better understand the nature of the interplay between this timescale separation and finite-size effects. As the size dependence is related to the disappearance of paused particles in the lattice, our method, which considers the lattice as being in two states (with or without paused particles), captures the relation between current and lattice size. To develop this, in the regime $k_p/\varepsilon$ and $k_u/\varepsilon \ll 1$ we assume the extreme cases of all particles regrouping in a single cluster, or relaxing to the standard TASEP dynamics when there are no paused particles. This single-cluster approximation reproduces well the simulated current-density relationships when the hypotheses are met (Figs.~\ref{fig:PBC_sim}-\ref{fig:heatmaps}). However, the relaxation process (necessary to achieve the steady-state current) being fast compared to the pausing/unpausing kinetics (see Fig.~\ref{fig:J_t}), we can generalise our approach to the multi-cluster regime, for which we approximating the current $J_0$ in the absence of paused particles by that of the standard TASEP Eq.~(\ref{eq:standard_TASEP}).

The single-cluster approximation is valid when the current is dominated by the state in which no paused particles are present. It is obviously destined to fail when this is not true, but then previously developed mean-field approaches are successful. 
Investigating the conditions when the approximation fails has provided insight into the intricate relationships between the dynamical parameters, the particle density and the lattice length.  On this basis this we have proposed a generalisation of our approach that provides good results in all the regions of the parameter space (see Fig.~\ref{fig:TASEP_pTASEP}). 

Finally, we present the extension of the single-cluster approximation to the open boundary conditions, showing how our two-state hypothesis can be extended to this situation. For instance, our model describes well how the current depends on the pausing rate $k_p$ across different orders of magnitudes.

A more systematic investigation of the open boundary case is out of the scope of this article. However, with this work we highlight how finite-size effects, often neglected in the literature of TASEP applied to transcription and translation, may be a major player to be taken into account when describing biological processes.

\begin{acknowledgments}
J.K. and L.V.D.Z. contributed equally to this work. J.K and L.V.D.Z. developed the theoretical framework; J.K. performed the simulations. L.C. conceptualized and supervised the work, supported by N.K. 
All authors discussed and interpreted the results and contributed to the final manuscript. 
This work was supported by the French National Research Agency (REF: ANR-21-CE45-0009).

\end{acknowledgments}

\appendix

\section{Simulations}
\label{app:sims}
We employ a standard Gillespie algorithm~\cite{gillespie1976general} for the simulations. The algorithm consists of two main phases. First, the system is allowed to run for a time $t_{ss}$ to reach steady state. After achieving steady state, data is collected for a duration $\Delta t$.

In the Gillespie algorithm, updates occur as Poisson processes, which means they are independent of each other. To ensure that the system reaches steady state and that the data collection period $\Delta t$ is sufficiently long, the total simulation time is made proportional to the lowest rate in the system. Unless otherwise stated,  we set $t_{ss} = 100 \times$ (lowest rate) and $\Delta t = 1000 \times$ (lowest rate). Thus, the total simulation time $t$ is equal $t=t_{ss} + \Delta t$.

\section{\label{app:current}Current and density profiles in the system approaching steady state}
In this Appendix we seek to compute the time evolution of the current produced by a periodic TASEP starting from a single cluster configuration, i.e all particles being adjacent one to the other. It has been proven \cite{Benassi1987} that the evolution of the density profile, seen as a continuous variable in time and space, in the TASEP is given by the unique entropic weak solution of a generalized Burgers' equation with appropriate boundary and initial conditions. From now onward, to avoid ambiguity, we will use $\overline{\rho}$ to refer to the number of particles divided by the lattice length, while for the density profile in the continuum limit we will use $\rho(x,t)$. Thus, in the case of a single cluster initial state and periodic boundaries on has: 
\begin{equation*}
    \begin{cases}
        \frac{\partial \rho(x,t)}{\partial t}-\varepsilon\left(1-2 \rho(x,t)\right) \frac{\partial \rho(x,t)}{\partial x}=0 & \forall t\geq 0, \ \forall x\in\left[0,L\right]\\
        \rho(x, 0)=1  & \forall x \in\left[0,\overline{\rho }L\right]\\ 
        \rho(x, 0)=0  & \forall x \in\left(\overline{\rho }L,L\right]\\ 
        \rho(0, t)=\rho(L, t) & \forall t\geq 0
    \end{cases},
\end{equation*}
where we have placed the beginning of the cluster on the first site for convenience.
\begin{figure*}[!t]
    \includegraphics[width = \textwidth]{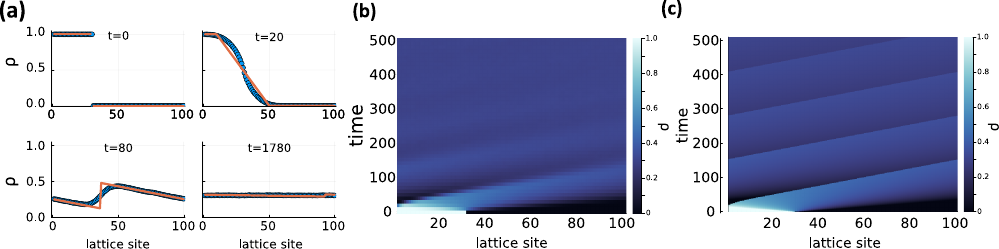}
    \caption{Cluster dissolution of the single-cluster configuration ($L=100$, $\rho = 0.3$). In (a), we show the density profile at different time points while the cluster dissolves. Panel (b) shows the numerical time evolution of the cluster and Panel (c) shows the time evolution of the single cluster according to the continuum limit approach.  \label{fig:burgers}
    }
\end{figure*}
\\
To lighten the notation it is convenient to adimensionalize the time variable trough the transformation $t'=t\varepsilon$, yielding:
\begin{equation}
    \begin{cases}
        \frac{\partial \rho(x,t)}{\partial t}-\left(1-2 \rho(x,t)\right) \frac{\partial \rho(x,t)}{\partial x}=0 & \forall t\geq 0, \ \forall x\in\left[0,L\right]\\
        \rho(x, 0)=1  & \forall x \in\left[0,\overline{\rho }L\right]\\ 
        \rho(x, 0)=0  & \forall x \in\left(\overline{\rho }L,L\right]\\ 
        \rho(0, t)=\rho(L, t) & \forall t\geq 0
    \end{cases}    
    \label{PDE_prob}
\end{equation}
where we have relabeled $t'$ as $t$.
Owing to the quasi linear nature of the Burgers' equation, the initial value contained at any given point $(x_0,0)$ remains constant along the characteristic lines:
\begin{equation*}
    x(t)=x_0+\left(1-2\rho(x_0,0) \right)t
\end{equation*}
up until the point in the space-time cylinder where two distinct characteristics meet. 
At that point, to obtain a weak solution of Eq.~(\ref{PDE_prob}), we consider a shock wave originating from that point and obeying the ODE:
\begin{equation}
    \Dot{x}(t)=1-\rho_l-\rho_r
    \label{shock}
\end{equation}

\begin{figure}[h]
    \includegraphics[width=0.8\columnwidth]{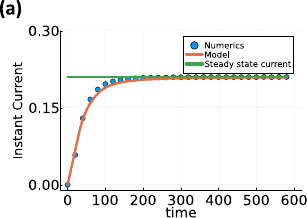}
    \caption{Comparison between the simulated and predicted current as a function of time, starting from a configuration with all particles clustered on the first sites. The horizontal line corresponds to the expected steady-state value. \label{fig:J_t}}
\end{figure}

that is, the time derivative of the shock profile is the mean of the time derivatives of the characteristics to its left and to its right \cite{Benassi1987}. 
When two characteristics diverge one from the other, and as such there exist points on the space-time cylinder which do not belong to the support of any characteristic line, there can be multiple weak solutions satisfying (\ref{PDE_prob}). 
The entropic one, which is the one actually describing the system, is obtained by connecting linearly (in space) the solutions existing at the boundaries of the divergence region, giving rise to what is often called a rarefaction fan.
From $t=0$ up until $t=\overline{\rho}L$ there will be a vertical shock separating a region with particle density $\rho=1$ from a region with $\rho = 0$ at position $x=0$, while a rarefaction fan will form at the point $x=\overline{\rho}L$. Thus we have 
\begin{equation*}
    \rho(x,t)=
    \begin{cases}
        1 & 0\leq x<\overline{\rho}L-t\\
        \frac{\overline{\rho}L+t-x}{2t} & \overline{\rho}L-t\leq x<\overline{\rho}L+t\\
        0 & \overline{\rho}L+t\leq x <L \,.
    \end{cases}
    ,  \
    t\in[0,\overline{\rho}L)
\end{equation*}
Starting from $t=\overline{\rho}L$ the $0-$density characteristics will intersect with the rarefaction fan's characteristics producing a shock, which will itself exist up until the last 0-characteristic will have intersected it. Depending on the starting cluster length the shock may or may not cross the system boundary, in which case its periodicity has to be accounted for. The shock curve $s(t)$ can be computed by solving the ODE (\ref{shock}) along with the condition $x(\overline{\rho}L)=0$, leading to:
\begin{equation*}
    s(t)=t-2\sqrt{\overline{\rho}L t}+\overline{\rho}L
    \, .
\end{equation*}
To understand where this equation is defined we have to compute the intersection point $x^*$ with $x=t-(1-\rho) L$ (the last $0-$characteristic) in non-periodic space and then check how many lattice lengths are spanned by the two curves before intersecting. We obtain $x^*=\left(\frac{1}{4\overline{\rho}}-1+\overline{\rho}\right)L $, as such the shock will cross the boundary $n=\left[\frac{x^*}{L}\right]=\left[\frac{1}{4\overline{\rho}}-1+\overline{\rho}\right]$ times, where the square brackets mean that we are considering the integer part of their argument. Thus we obtain:
\begin{equation*}
    \rho(x,t)=
    \begin{cases}
        0 & 0\leq x<s(t)\\
        \frac{\overline{\rho}L+t-x}{2t} & s(t)\leq x<t+\overline{\rho}L\\
        0 & t+\overline{\rho }L\leq x <L
    \end{cases}
\end{equation*}
for $\forall t\in\left[\overline{\rho}L,(1-\overline{\rho})L\right)$, while for $1=1,..,n$ we have
\begin{equation*}
    \rho(x,t)=
    \begin{cases}
        0 & 0\leq x<s(t)-iL\\
        \frac{(\overline{\rho}-i)L+t-x}{2t} & s(t)-iL\leq x<t+(\overline{\rho}-i)L
        \\
        0 & t+(\overline{\rho}-i)L\leq x <s(t)-(i-1)L\\
        \frac{(\overline{\rho}-i+1)L+t-x}{2t} & s(t)-(i-1)L\leq x <L
    \end{cases}
\end{equation*} 
for $t\in\left[(1-\overline{\rho})L+(i-1)L,(1-\overline{\rho})L+iL\right)$
. Finally, we get to the time instant $t^*=\frac{L}{4\overline{\rho}}$ where the last $0-$characteristic line meets the shock:
\begin{equation*}
    \rho(x,t)=
    \begin{cases}
        \frac{(\overline{\rho}-n-1)L+t-x}{2t} & 0\leq x<t-nL-(1-\overline{\rho})L \\
        0 &  t-nL-(1-\overline{\rho})L\leq x<s(t-nL)\\
        \frac{(\overline{\rho}-n)L+t-x}{2t} & s(t-nL)\leq x <L
    \end{cases}
\end{equation*}
for $t\in \left[(1-\overline{\rho})L+nL,\frac{L}{4\overline{\rho}}\right)$.

Some of the characteristics will cross the lattice to intercept other characteristics from the same rarefaction fan, leading to a new shock $z(t)$, which is going to propagate to infinity. Solving (\ref{shock}) we obtain:
\begin{equation*}
    z(t)=\left(\overline{\rho}-n-\frac{1}{2}\right)L+ct
\end{equation*}
and imposing the condition $z(\frac{L}{4\rho})=x^*-nL=$ we get:
\begin{equation*}
    c=1-2\rho.
\end{equation*}
The shock will meet the boundary at time:
\begin{equation*}
   t^{**}=\frac{L}{4\overline{\rho}}+\frac{L-(x^*-nL)}{c} 
\end{equation*}
so we have
\begin{equation*}
    \rho(x,t)=
    \begin{cases}
        \frac{(\overline{\rho}-n-1)L+t-x}{2t} & 0\leq x<z(t)\\
        \frac{(\overline{\rho}-n)L+t-x}{2t} & z(t)\leq x <L
    \end{cases}
\end{equation*}
for $t\in \left[\frac{L}{4\overline{\rho}},t^{**}\right)$.
From $t^{**}$ onwards the system will repeat periodically up to infinity, with a time period $\tau$ given by:
\begin{equation*}
    \tau=\frac{L}{c}=\frac{L}{1-2\rho} \,.
\end{equation*}
Thus $\forall t\geq t^{**}$ one has:
\begin{equation*}
    \rho(x,t)=
    \begin{cases}
        \frac{(\overline{\rho}-n-m-2)L+t-x}{2t} & 0\leq x<c(t-t^{**})-mL\\
        \frac{(\overline{\rho}-n-m)L+t-x}{2t} & c(t-t^{**})-mL\leq x <L
    \end{cases}
\end{equation*}
where $m=\left[\frac{t-t^{**}}{\tau}\right]$. Now all one needs to obtain Eq.(\ref{eq:J0}) is to transform back the time unit to the two integrals.

In Fig.~\ref{fig:burgers} we compare the predictions for the evolution of the density profile compared to simulations. In panel (a) we show how the method of characteristics provides a good approximation for the evolution of the density profile starting from the situation at $t=0$ where all particles are clustered on the first sites. Panels (b) and (c) show the continuous evolution of the density profile in time and space for the simulated system (b) and for the analytical approach used in this appendix (c).

Figure~\ref{fig:J_t} shows how the current $J_0$, computed in Eq.~(\ref{eq:J0}) after plugging $\rho(x,t)$ obtained in this Appendix into Eq.~(\ref{current_integral}). We notice that the TASEP steady-state value given in Eq.~(\ref{eq:standard_TASEP}) is reached in a short time compared to the duration of the state with zero paused particles given by $k_p \rho L$. This further justifies approximating the current in this state with Eq.~(\ref{eq:standard_TASEP}), as done in the multi-cluster approach (Section~\ref{sec:multi-cluster}).

\bibliography{ReferencesPausedTASEP}

\end{document}